# The State of Open Data in Latvia: 2014

Uldis BOJĀRS[1] and Renārs LIEPIŅŠ

Institute of Mathematics and Computer Science, University of Latvia,
Raina bulvaris 29, Riga, LV-1459, Latvia

uldis.bojars@gmail.com, renars.liepins@lumii.lv

**Abstract.** This paper examines the state of Open Data in Latvia at the middle of 2014. The study is divided into two parts: (i) a survey of open data situation and (ii) an overview of available open data sets. The first part examines the general open data climate in Latvia according to the guidelines of the OKFN Open Data Index making the results comparable to those of other participants of this index. The second part examines datasets made available on the Latvia Open Data community catalogue, the only open data catalogue available in Latvia at the moment. We conclude that Latvia public sector open data mostly fulfil the basic criteria (e.g., data is available) of the Open Data Index but fail on more advanced criteria: the majority of data considered in the study are not published in machine-readable form, are not available for bulk download and none of the data sources have open license statements.

## 1. Introduction

The data openness, and open government data in particular, is gaining traction across the world. It has been particularly advocated for government data, because it supports government action transparency on the one hand and facilitates public data reuse on the other, and results in better service for citizens and companies (Davies, 2010). The reuse of public data has a great economic potential, which was emphasized by Neelie Kroes, the Vice President of the EU and Commissioner for the Digital Agenda, when she said: "Data is a 21$^{st}$ century commodity: it's the new oil. There's almost no limit to the economic and social wonders it can generate [...]."[2]

The open government data movement was pioneered by the US and the UK with data.gov and data.gov.uk initiatives. In Europe this movement has been encouraged by the Public Sector Information (PSI) directive on the reuse of public sector information (European Commission, 2003). The directive was first released in 2003 and has been revised last year (2013) to focus on issues related to publication and reusability of public datasets as open data.

According to the Open Knowledge Definition, a dataset is Open Data if it "can be freely used, reused and redistributed by anyone – subject only, at most, to the requirement to attribute and share-alike" (Open Definition, 2009). In order for this to work, it must be open both legally and technically. The legal openness is ensured by publishing data under an open licence. The technical openness is achieved by publishing data in machine-readable formats (Open Knowledge Foundation, 2012) and ensuring the discoverability of the datasets themselves[3].

---

[1] Corresponding Author.
[2] http://blog.okfn.org/2012/09/20/rest-assured-the-eu-is-behind-you-says-european-commissioner-neelie-kroes-to-okfestival-participants/

Although there has been much progress towards open government data worldwide, there are huge differences in the state of open data in different countries. According to the Open Data Census that surveys the worldwide state of data openness, some countries (UK, US) have very high scores (more than 900 out of 1000) while most are below 500 and some as low as 85 (Open Knowledge Foundation, 2013). There are still many countries not covered by the index, which means the state of open data in them is not known.

The open data initiative is just starting in Latvia at the government level and Latvia currently is not included in the Open Data Census. The goal of this study is to survey the current open data situation in Latvia and to create a reference point that can be used to assess future progress.

The paper is structured as follows. It starts with a survey of related work, followed by an overview of the methodology used to assess the state of open data in Latvia. We use two independent assessment strategies the results of which are presented after the methodology section. First, we survey the overall open data "climate" following the approach of the Open Data Index. In the second part of the study we examine the actual datasets available on the data.opendata.lv catalogue. We conclude the paper with a summary and a discussion of the results.

## 2. Related Work

Public sector open data are studied in the existing research from multiple perspectives: the infrastructure needed for provisioning and reuse of open data (Zuiderwijk et al., 2013), applying Linked Data principles for improved publishing, linking and exploring of open government data (Ding et al., 2011; Shadbolt et al., 2012), exploring the state of international open data catalogues (Martin et al., 2013) or examining the impact of EU PSI directives on government opening up their data (Janssen, 2011). A part of this research looks into better infrastructures and richer ways of publishing, linking, transforming and reusing existing datasets. Our study indicates that Latvia is in earlier stages of open government data and, before such techniques can be applied, data needs to be open and available as machine-readable data on the Web.

Our goal is to explore and record the state of open data in Latvia, which the future progress can be evaluated against. Topics that are related to ours are studies of the state of country open data and studies of criteria that open datasets can be evaluated on. Surveys covering the Baltic States would be particularly useful in that they can provide a local context that Latvia's results can be compared to, but we were not able to find existing publications that survey the state of open data in Latvia or the Baltic States in general. A study of data catalogue quality criteria described in Kučera et al. (2013) mentions an example of the Czech open data catalogue but it does not explore the state open data in the country in further detail.

---

[3] http://www.w3.org/TR/vocab-dcat/

Existing studies that cover EU countries include the Open Data Index (Open Knowledge Foundation, 2013) and the EU Public Sector Information Scoreboard (ePSI platform, 2014). The Open Data Index is an effort by the Open Knowledge Foundation (OKFN)[4] where a community of open data advocates surveyed 10 types of datasets that capture a range of key national information and are classified as "high value" datasets in the G8 Open Data Charter (UK Cabinet Office, 2013). Each dataset of Open Data Index is evaluated according to 9 weighted criteria (ranging from "does the data exist at all?" to data machine-readability and availability for bulk downloads) and 3 descriptive criteria such as the URL of data available online[5]. Each year the results are reviewed by a panel of experts and presented as the Open Data Census that shows the state of open data across participating countries. The census for 2013 contains 70 countries and is available at https://index.okfn.org/country/ however it does not include data about Latvia (Open Knowledge Foundation, 2013).

A study a large data catalogue described in Martin et al. (2013) surveys the PublicData.EU catalogue that aggregates data from other EU open data catalogues. Datasets are examined according to the "5-star" classification of linked open data proposed by Tim Berners-Lee (2009). The survey has similarities to our study of data.opendata.lv but its methodology is not particularly useful in our case because the open data currently available in Latvia do not even have open data licences (thus getting 0 out of 5 stars in the "5-star" classification). It would be useful to examine Latvia open data landscape according to these criteria when the situation improves – when there is a larger amount of diverse data available and when open data are assigned proper licences.

The EU Public Sector Information Scoreboard[6] evaluates the state of data openness. The scoreboard is available on the Web and describes itself as a crowdsourced tool to measure the status of Open Data and PSI re-use throughout the EU (ePSI platform, 2014). The scoreboard assesses the state of the Open Data at the policy level, i.e. how well the governments comply with the EU PSI directive from 2003. This policy-level view is different from the dataset-centric approach of our study in the sense that a government may have fully implemented the EU PSI directive 2003 and received an acceptable PSI scoreboard valuation yet that does not guarantee that there are open and reusable datasets made available by public sector institutions.

Our study focuses on the availability of open data sets and on open data in the sense of PSI directive from 2013 where much more emphasis is placed on machine-readability and open licences[7]. Thus we will use the Open Data Index approach as a template for the rest of the paper.

---

[4] http://okfn.org
[5] https://index.okfn.org/about/#criteria
[6] http://www.epsiplatform.eu/content/psi-scoreboard-indicator-list
[7] http://ec.europa.eu/digital-agenda/en/news/what-changes-does-revised-psi-directive-bring

## 3. Methodology

Latvia does not have an official open data portal therefore we have to look for other sources of information. Our approach is two-fold:
1. survey the overall data openness "climate" by analysing Latvia public sector information using an approach of the Open Data Index, making the results comparable to other countries included in the index;
2. examine the datasets available on the Latvia Open Data community catalogue (data.opendata.lv).

This two-fold approach is necessary because the first part of the analysis alone would only characterize the level of "data openness" of government institutions but would not tell us what is the state of open, machine-readable datasets apart from the types of data considered by the Open Data Index.

The first part of the study was performed by examining selected data types and following the Open Data Index regarding the choice of types of data, evaluation criteria (e.g. if the dataset is available for bulk download) and dataset ranking principles. For each of the 10 types of data we examined if this information is available and how it satisfies the evaluation criteria.

The results of the first part of the study show us that while for most types of data included in the survey the information is available online in some form, none of these types of data are available for bulk download or have open data licenses. These results may characterize the overall PSI data openness in Latvia but they do not tell us what machine-readable PSI datasets are available, if any.

In the second part of the study we examine data.opendata.lv catalogue developed by the Latvia Open Data community. At the time of writing Latvian government did not have an open data portal or a website providing information about open data in Latvia. While some government data could be found online, it is scattered over the web, is not easily discoverable and is not in a readily usable form. By analysing the community catalogue we can see what open datasets are available and what types of data are considered important by the community. In order to understand the dynamics of the catalogue we also look at the relation between dataset creation and open data community activities such as hackathons.

By combining both parts of the methodology we get an overview of both the overall data openness "climate" in Latvia and the state of open, machine-readable datasets ready for reuse.

## 4. Assessment of Latvian Public Sector Information

The Open Data Index, which we use as a foundation for this assessment, is focused on 10 datasets that capture a range of key national information (Open Knowledge Foundation, 2013) and are classified as "high value" datasets in the G8 Open Data Charter (UK Cabinet Office, 2013).

The list of datasets together with the gathered answers to questions that capture the state of availability and openness of the given dataset are shown in the Table 1. The score is calculated according to the criteria and weighting rules given in the Open Data Index, i.e. it is a weighted sum of the scores for each[8]. Criteria score is shown in parenthesis after the corresponding question on the top row in the Table 1. The maximum possible score for each dataset is 100.

---
[8] https://index.okfn.org/about/#criteria

The total score (a sum of dataset scores) for Latvia is 440. That would place Latvia in the 27th-29th place (out of 70) in the Open Data Census 2013 (Open Knowledge Foundation, 2013). Lithuania is in the 50th place with a total score of 320[9], while Estonia is currently not included in the Census. In terms of the global ranking where the UK has the best score of 940 (out of 1000) there is a place for significant improvement for the Baltic States.

Out of the 10 data types examined the worst performing is the government spending data where no detailed level information is publicly available[10], and the national map and company register which are not free-of-charge and thus users may incur significant costs if they need to work with this data.

|  | Data exists (5) | It is digital (5) | Publicly available (5) | Free of charge (15) | It is online (5) | Machine readable (15) | Available in bulk (10) | Open License (30) | Up-to-date (10) | Score |
|---|---|---|---|---|---|---|---|---|---|---|
| Transport Timetables | Y | Y | Y | Y | Y | N | N | N | Y | 45 |
| Government Budget | Y | Y | Y | Y | Y | N | Y | N | Y | 55 |
| Government Spending | Y | Y | N | N | N | - | - | - | Y | 20 |
| Election Results | Y | Y | Y | Y | Y | N | N | N | Y | 45 |
| Company Register | Y | Y | Y | Y | Y | Y | Y | N | Y | 70 |
| National Map | Y | Y | Y | N | N | Y | N | N | ? | 30 |
| National Statistics | Y | Y | Y | Y | Y | Y | N | N | Y | 60 |
| Legislation | Y | Y | Y | Y | Y | N | N | N | Y | 45 |
| Postcodes / Zipcodes | ? | ? | N | N | N | N | N | N | ? | 0 |
| Emissions of pollutants | Y | Y | Y | Y | Y | Y | Y | N | Y | 70 |
|  |  |  |  |  |  |  |  |  |  | 440 |

**Table 1.** Summary of the Open Data Index survey for Latvia.

In order to get an overview of the strong and weak points in the state of government data openness in Latvia we aggregated the results by evaluation criteria. The graphical representation is shown in Figure 1. As can be seen, the situation with existence of digital data is generally very good (the first three columns are almost completely green) and the data is mostly publicly available and accessible online (8 out of the 10 datasets are publicly available). Thus we can conclude that the recommendations of the European Union PSI directive from the 2003 are mostly realized in practice.

---

[9] https://index.okfn.org/country/overview/Lithuania/
[10] According to the criteria of Open Data Index: "Records of actual (past) national government spending at a detailed transactional level; at the level of month to month government expenditure on specific items (usually this means individual records of spending amounts under $1m or even under $100k)."

The next three criteria (machine readability, bulk access, and, most importantly, an open licence), which are very important for being able to use the data, show much worse results – only a few datasets are machine-readable and almost none are available for bulk download.

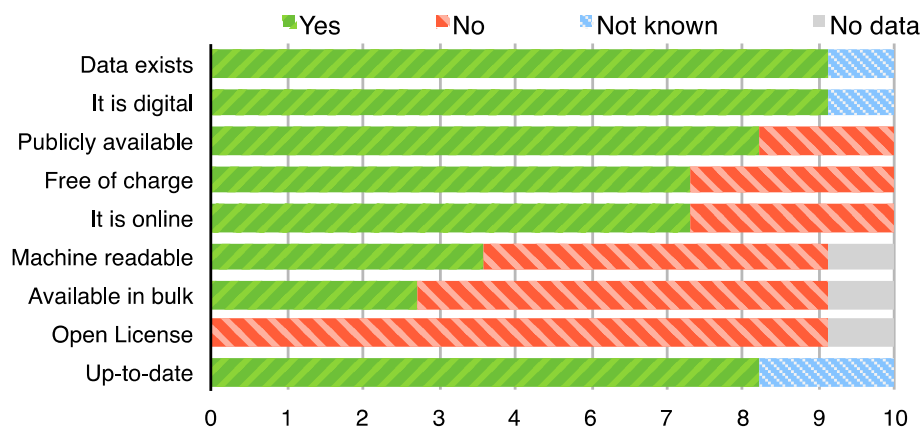

**Figure 1.** Distribution of answers for each criteria from the Open Data Index for Latvian datasets.

An important consideration is that none of the types of data examined have an open license attached to them thus the users may not know if and how they may use and reuse this information. Thus, according to the OKFN Open Data Handbook (Open Knowledge Foundation, 2012) none of these datasets should actually be considered Open Data. The importance of having an open licence is reflected by the weights of the Open Data Index criteria where the weight for having an open licence attached is 30% of the total. This, however, does not fully prevent people from doing something with this data because this is public sector information and it may have more open conditions for use (defined in laws and regulations) than private sector information that would be protected by copyright.

## 5. Survey of data.opendata.lv catalogue

The second part of the study examines the available open government datasets and the activity of the open data community in Latvia. In particular, we examine the open data catalogue[11] created by the Latvia Open Data community. To the best of our knowledge it is the only open data catalogue in Latvia at the moment.

This source contains open data that are available in machine-readable formats and for bulk download. It allows users to find ready-to-use datasets without a need for complex and time-consuming data gathering and transformation tasks often required when collecting government data in the form they are currently published in (e.g. as HTML pages). The catalogue allows us to examine both what open datasets are available in Latvia and what types of data are deemed important enough by the community in order to do the tasks of data collecting, transformation and adding to the catalogue.

---

[11] http://data.opendata.lv

As of January 2014 the data.opendata.lv portal contained 25 datasets added between May 2012 and December 2013. Next we will look at the types of datasets in the catalogue and at the relation between dataset creation time and the overall activity of the open data community.

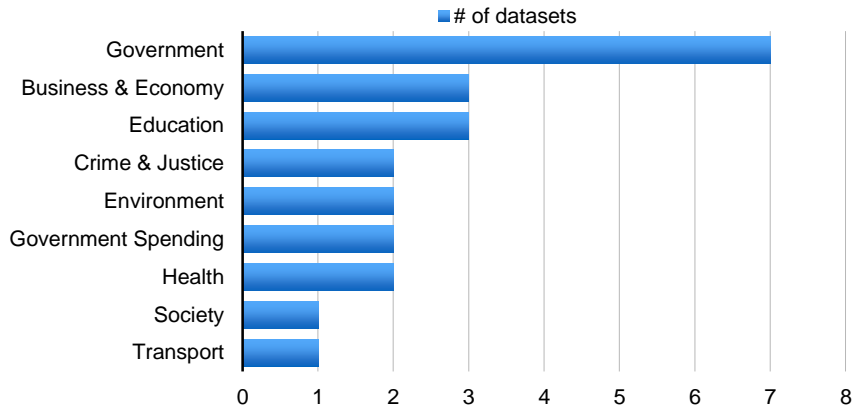

**Figure 2.** Types of datasets in the data.opendata.lv catalogue.

Figure 2 shows the types of datasets in the catalogue, classified according to the high-level categories used in the UK Open Data portal[12]. The top three types of data in our catalogue are Government (7 datasets), Business & Economy (3 datasets) and Education (3 datasets). The datasets differ in size from tables with tens of rows for statistical data to databases with hundreds of thousands of entries.

An example of a Government type of datasets is the detailed voting information of MPs of Saeima (Parliament of the Republic of Latvia). The dataset for the 10th Saeima consists of a list of MPs and records of every vote made by them (except for closed voting sessions), resulting in more than 157 thousand entries[13]. This dataset was collected by crawling the original voting data represented as separate HTML tables[14] and transforming it into data records representing MPs, the topics that were voted on and the votes cast. This information was later used for analysing parliamentary voting patterns and networks (Bojārs et al., 2012).

The dynamics of dataset creation, shown in Figure 3, indicates that there are activity peaks and quiet periods between them. The peaks of activity, with three or more datasets added per month, are May 2012, June 2012 and December 2013.

These activity peaks are related to open data hackathons[15] – face-to-face meetings organized by Latvia Open Data community where groups of volunteers come together and aim to do something useful with open data. Often the participants would find some publicly available data they are interested in, transform the data into an easy-to-use form and submit it to data.opendata.lv.

---

[12] http://data.gov.uk
[13] http://data.opendata.lv/jbaiza/10-saeimas-balsojumi
[14] http://titania.saeima.lv/LIVS10/SaeimaLIVS2_DK.nsf/DK?ReadForm&calendar=1
[15] "A hackathon (also known as a hack day, hackfest or codefest) is an event in which computer programmers and others involved in software development collaborate intensively on software projects." – http://en.wikipedia.org/wiki/Hackathon

The open data community and hackathons are an important catalyst of open data in Latvia, resulting in the first open datasets becoming available. The 1st Latvia Open Data Hackaton[16] took place in December 2011; the 2nd hackathon[17] took place in June 2012 and the third hackathon – in December 2013.

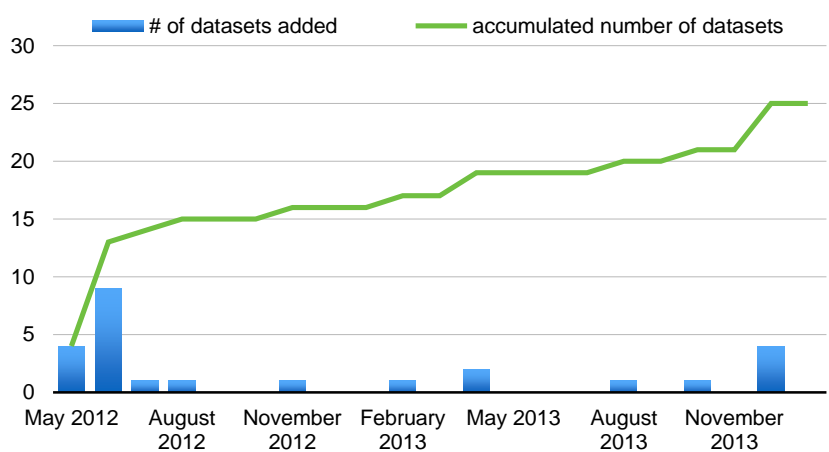

**Figure 3.** Aggregate dataset count (green line) and number of dataset added each month (blue bars) by the Latvian open data community.

The timing of the last two hackathons coincides with the peaks of activity when 9 datasets were added in June 2012 and 4 datasets - in December 2013. The timing of the first activity peak is related to both the creation of this catalogue and the 1st Latvia Open Data Hackathon that generated two of the 4 datasets added to the catalogue in May 2012 - the results of the recent parliamentary elections and a database listing donations to political parties in Latvia.

While the work done by volunteers is valuable it cannot act as a substitute for government open data activities and should rather complement them. The limitations of volunteer open data activities are that they are spontaneous and ad-hoc, which means that data can be created quickly but there are no guarantee that the datasets will be maintained and kept up-to-date. The government institutions that hold the original data are in better position to create the same datasets with less effort, with clear open data licenses and to ensure that the information is kept up-to-date.

---

[16] http://opendata.lv/2011/12/10/pirmais-open-data-day-hackathon-ir-noticis
[17] http://opendata.lv/2012/06/17/otrais-atverto-datu-hakatons/

## 6. Conclusions

In this study we gave a general assessment of the state of open data in Latvia and examined the open datasets available as of 2014.

The first part of the study followed the guidelines of the Open Data Index and focused on 10 types of high-value government data. The results are comparable to those of other countries included in the index and would place Latvia in the $27^{th}$–$29^{th}$ place in the Open Data Census 2013. Most types of data studied are freely available on the web in some form. The weak points in terms of types of open data in Latvia are the government spending data (no detailed level information is available) and the national map is publicly available but not free-of-charge.

The summary of the first part of the study shows that Latvia mostly fulfils the basic criteria and that the data is available online. However, it fails on important criteria that make the data usable – the majority of data considered in the study are not published in machine-readable form, is not available for bulk download and none of the data sources have open license statements. We conclude that there is much potential for further improvements in opening up the data and unlocking the value of public sector data in Latvia.

In the second part of the study we examined the datasets in the data.opendata.lv catalogue that was the only open data catalogue in Latvia at the time of writing. It contains datasets gathered by the Latvian Open Data enthusiast community. They were better than the data we looked at in the first part of the study in the sense that they were more "actionable" (available in a machine-readable form and for bulk download). In total there were 25 datasets about varied topics, mostly government data. When examining the creation dates of the datasets we noted a strong correlation between the dataset creation dates and the timing of open data hackathon events. This leads us to believe that these datasets are not produced systematically but are a result of ad-hoc group rallies.

These datasets are a good starting point for open data consumers because they are readily findable, are easily usable and may act as examples for others who look for how to open their data. However, they have two major drawbacks. First, most of these datasets are not formally "open" according to the Open Definition because the community does not own the data (they have collected it from government websites) and thus cannot give permissions and add open licences that were not present in the original source. The second problem is that the data may not be up-to-date because of the ad-hoc nature of data gathering efforts.

Thus we think that the open data community is a valuable resource that could help the government initiative by converting data to more formats, but at least open licences and the information being up-to-date must come from systematic government efforts. Data discoverability also needs to be considered by the original data sources so that catalogues can be created and updated automatically, based on metadata associated with the open data sets.

## Acknowledgements

We acknowledge ESF project 013/0005/1DP/1.1.1.2.0/13/APIA/VIAA/049 for providing a partial support for this study.

# References

Berners-Lee, T. (2009). Linked Data - Design Issues. Revision: 18.06.2009. URL: http://www.w3.org/DesignIssues/LinkedData.html

Bojārs, U., Ručevskis, P. and Krebs, V. (2012). Exploring the Networks in Open Public Data. In *Using Open Data: policy modeling, citizen empowerment, data journalism workshop*. Brussels, Belgium, June 2012. URL: http://www.w3.org/2012/06/pmod/exploring-open-data-networks.html

Davies, T. (2010). Open data, democracy and public sector reform: a look at open government data use from data.gov.uk. *MSc thesis at Oxford Internet Institute*. URL: http://www.opendataimpacts.net/report/

Ding, Li et al. (2011). TWC LOGD: A portal for linked open government data ecosystems. *Web Semantics: Science, Services and Agents on the World Wide Web*, 9.3, 325-333.

ePSI platform (2014). EU Public Sector Information Scoreboard. URL: http://www.epsiplatform.eu/content/european-psi-scoreboard.

European Commission (2003). Directive 2003/98/EC of the European Parliament and of the Council of 17 November 2003 on the Re-use of Public Sector Information (2003). *O.J. L345/90*. URL: http://eur-lex.europa.eu/LexUriServ/LexUriServ.do?uri=OJ:L:2003:345:0090:0096:EN:PDF.

Janssen, K. (2011). The influence of the PSI directive on open government data: An overview of recent developments. *Government Information Quarterly*, 28.4, pp. 446-456.

Kučera, J., Chlapek, D. and Nečaský, M. (2013). Open Government Data Catalogs: Current Approaches and Quality Perspective. In *Technology-Enabled Innovation for Democracy, Government and Governance*. Springer Berlin Heidelberg, pp. 152-166.

Martin, S., Foulonneau, M. and Turki, S. (2013). 1-5 Stars: Metadata on the Openness Level of Open Data Sets in Europe. In *Metadata and Semantics Research*. Springer International Publishing, pp. 234-245.

Open Definition (2009), version 1.1. URL: http://opendefinition.org/od/

Open Knowledge Foundation (2012). The Open Data Handbook. URL: http://opendatahandbook.org/pdf/OpenDataHandbook.pdf

Open Knowledge Foundation (2013). The Open Data Index. 28.10.2013. URL: https://index.okfn.org/country

Shadbolt, N., O'Hara, K., Berners-Lee, T., Gibbins, N., Glaser, H. and Hall, W. (2012). Linked open government data: Lessons from data.gov.uk. *IEEE Intelligent Systems*, 27.3, pp. 16-24.

UK Cabinet Office (2013). G8 Open Data Charter and Technical Annex, 18.06.2013. https://www.gov.uk/government/publications/open-data-charter/g8-open-data-charter-and-technical-annex

Zuiderwijk, A., Janssen, M. and Jeffery, K. (2013). Towards an e-infrastructure to support the provision and use of open data. *Conference for E-Democracy and Open Governement*, *2013*.

## Authors' Information

**Uldis Bojārs**

Dr. Uldis Bojārs is a Researcher at the Institute of Mathematics and Computer Science, University of Latvia. His areas of interest include Open Data, Linked Data and Social Semantic Web. He took part in the W3C Semantic Web Education and Outreach Interest Group and the W3C Library Linked Data Incubator Group, and was a Co-Chair of the Social Data on the Web (SDoW) workshop series. Uldis is a co-founder of the SIOC Project aimed at applying Semantic Web technologies to Social Web sites. He has a PhD in Computer Science from the National University of Ireland, Galway.

**Renārs Liepiņš**

Renārs Liepiņš is a Researcher at the Institute of Mathematics and Computer Science, University of Latvia. His areas of interest include Domain Specific Graphical Languages and Tools, Semantic Web, Linked Data and Open Data.